\documentclass[galaxies,article,accept,pdftex,moreauthors]{Definitions/mdpi} 

\usepackage[defaultcolor=red]{changes}

\makeatletter
\@namedef{Changes@AuthorColor}{red}
\colorlet{Changes@Color}{red}
\makeatother

\firstpage{1} 
\pubvolume{1}
\issuenum{1}
\articlenumber{0}
\pubyear{2022}
\copyrightyear{2022}
\externaleditor{Academic Editor: {Firstname Lastname}}
\datereceived{} 
\daterevised{} 
\dateaccepted{} 
\datepublished{} 
\hreflink{https://doi.org/} 

\newcommand{\aap}{Astron. Astrophys.}

\newcommand{\apjl}{Astrophys. J.}
\newcommand{\apjs}{Astrophys. J. Suppl.}

\newcommand{\pasp}{Publ. Astron. Soc. Pacific}
\def\sii{[S\,{\sc ii}]}
\def\ha{H$\alpha$}
\def\kms{\relax \ifmmode {\,\rm km\,s}^{-1}\else \,km\,s$^{-1}$\fi}

\Title{Large-Scale Ejecta of Z CMa---Proper Motion Study and New Features Discovered}
\TitleCitation{Large-Scale Ejecta of Z CMa---Proper Motion Study and New Features Discovered}

\Author{Tiina Liimets 
$^{1,*}$\orcidA{}, Michaela Kraus~$^{1}$\orcidB{}, Lydia Cidale~$^{2,3}$\orcidC{}, Sergey Karpov~$^{4}$\orcidD{} and Anthony Marston~$^{5}$\orcidE{}}

\AuthorNames{Tiina Liimets, Michaela Kraus, Lydia Cidale, Sergey Karpov, Anthony Marston}

\AuthorCitation{Liimets, T.; Kraus, M.; Cidale, L.; Karpov, S.; Marston, A.}

\address{%
$^{1}$ \quad Astronomical Institute, Czech Academy of Sciences, Fri\v{c}ova 298, 25165 Ond\v{r}ejov, Czech Republic; michaela.kraus@asu.cas.cz\\
$^{2}$ \quad Instituto de Astrof\'isica de La Plata (CCT La Plata-CONICET, UNLP), Paseo del Bosque S/N, La Plata, B1900FWA, Buenos Aires, 
  Argentina; lydia@fcaglp.fcaglp.unlp.edu.ar\\
$^{3}$ \quad Departamento de Espectroscop\'ia, Facultad de Ciencias Astron\'omicas y Geof\'isicas, Universidad Nacional de La Plata, Paseo del Bosque S/N, La Plata, B1900FWA, Buenos Aires, Argentina\\
$^{4}$ \quad Institute of Physics of the Czech Academy of Sciences (FZU AV \v{C}R), Na Slovance 2, \mbox{Praha {8}, 18200 Prague, Czech Republic}; karpov@fzu.cz\\
$^{5}$ \quad European Space Agency, European Space Astronomy Centre, Camino Bajo del Castillo, s/n Urbanizaci\'on Villafranca del Castillo Villa\~nueva de la Ca\~nada, E-28692 Madrid, Spain; anthony.marston@esa.int}

\corres{Correspondence: tiina.liimets@asu.cas.cz} 

\abstract{Z Canis Majoris is a fascinating early-type binary with a Herbig Be primary 
and a FU Orionis-type secondary. Both of the stars exhibit sub-arcsecond jet-like ejecta.  
In addition, the primary is associated with the extended jet as well as with the 
large-scale outflow. In this study, we investigate further the nature of the {large-scale} 
outflow, which has not been studied since its discovery almost three and a half 
decades ago. We present proper motion measurements of individual features 
of the large-scale outflow and determine their kinematical ages. Furthermore, 
with our newly acquired deep images, we have discovered additional faint arc-shaped 
features that can be associated with the central binary.}

\begin{document}





\section{Introduction} \label{sec:intro}

Z Canis Majoris (Z CMa) has intrigued astronomers for decades. It is an active early-type 
emission line binary consisting of a Herbig Be primary and a FU Orionis-type (FU Ori) 
secondary separated by {$0''.1$} (e.g., \citet{2017A&A...597A..91B}). Current high-spatial 
resolution observations show that the primary is located in the northwest (NW) direction from the secondary 
(e.g., \citet{2017A&A...597A..91B}; \citet{2022NatAs...6..331D}).
The light curve of the system is rich in long-term variability, months to years, as well as in day-by-day variability 
with a non-periodic nature and varying amplitudes (\citet{2020A&A...643A..29S} and references therein). 

\begin{figure}[H]
\begin{centering}
\includegraphics[width=9.5cm, trim=0 0 0 0]{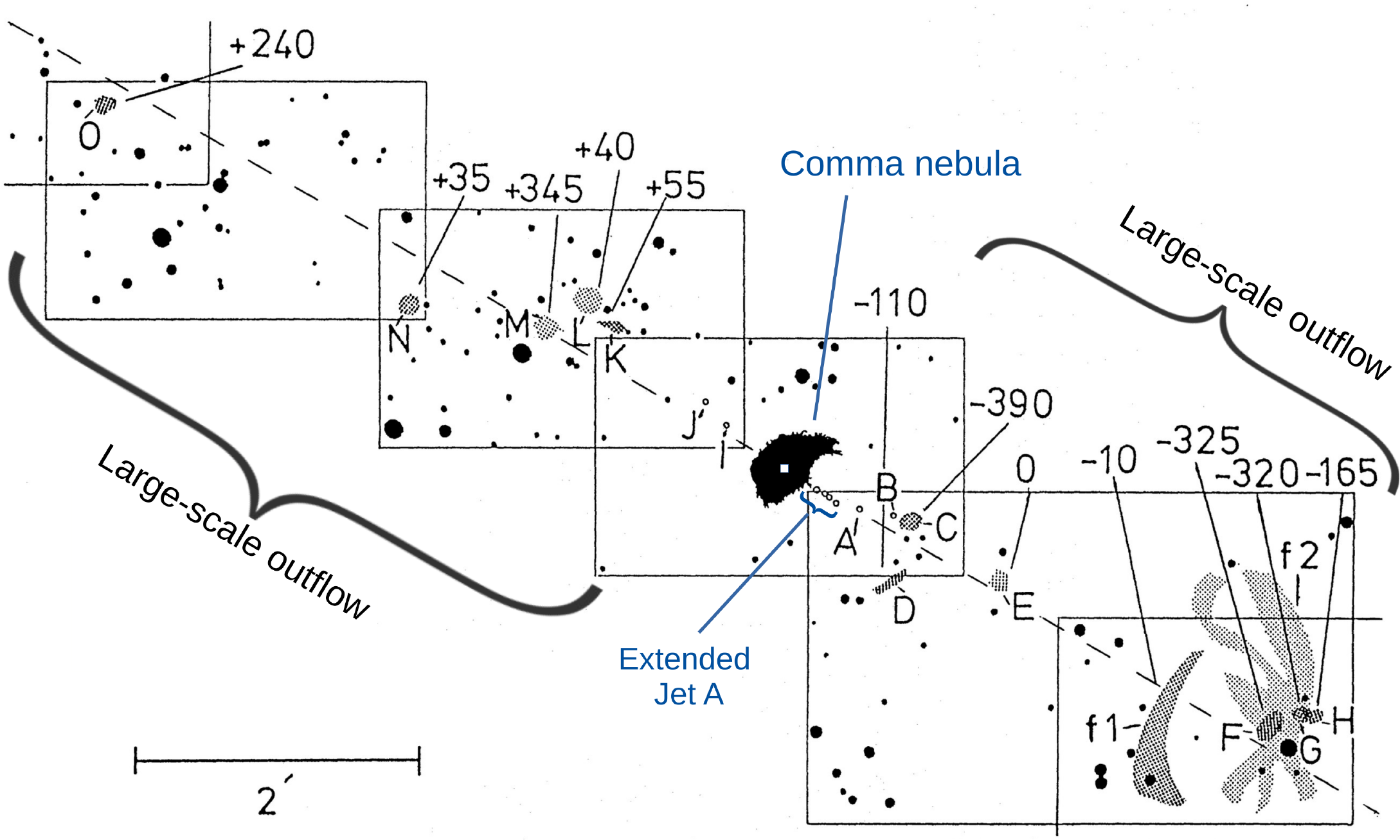}
\includegraphics[width=3.3cm, trim=0 0 0 0]{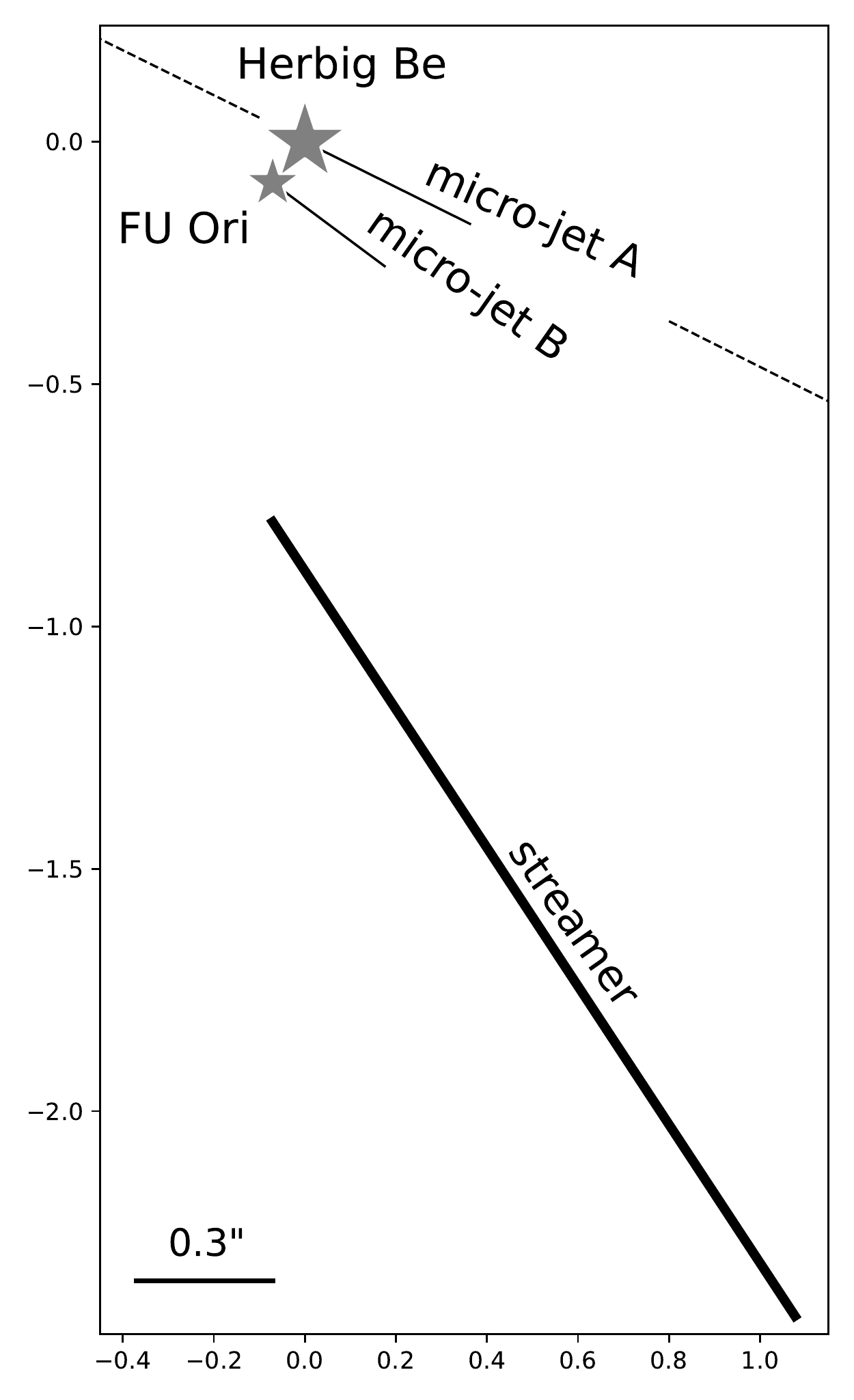}
\caption{{{ (\textbf{Left}): } Schematic view of the various extended nebular features surrounding Z CMa. 
The stars in the field of view are drawn as black filled circles, large-scale outflow features are marked 
with grayish areas or circles with black contures. The dashed line 
is drawn at the 60\textdegree\ position angle. Individual numbers refer to radial 
velocities (\citet{1989A&A...224L..13P}). Field of view is $9'.5\times5'.6$. 
The base of the figure is depicted from \citet{1989A&A...224L..13P}. Reproduced with permission \copyright~ESO. 
{ (\textbf{Right}):} Schematic view of the sub-arcsecond features around Z CMa presented in true proportions. 
The lengths of the micro-jets 
are taken from \citet{2010ApJ...720L.119W} and those for the streamer are taken from \citet{2022NatAs...6..331D}. 
The dashed line represents the position angle of the large-scale outflow.
Field of view is  $1''.6\times2''.7$. A white square representing the same size is drawn on the left panel at the 
position of the central binary inside the comma nebula. On both panels, north is up and east is to the left. }}
\label{F-poetzel}
\end{centering}
\end{figure}

\begin{figure}[H]
\begin{centering}
\includegraphics[width=13.5cm, trim=0 0 0 0]{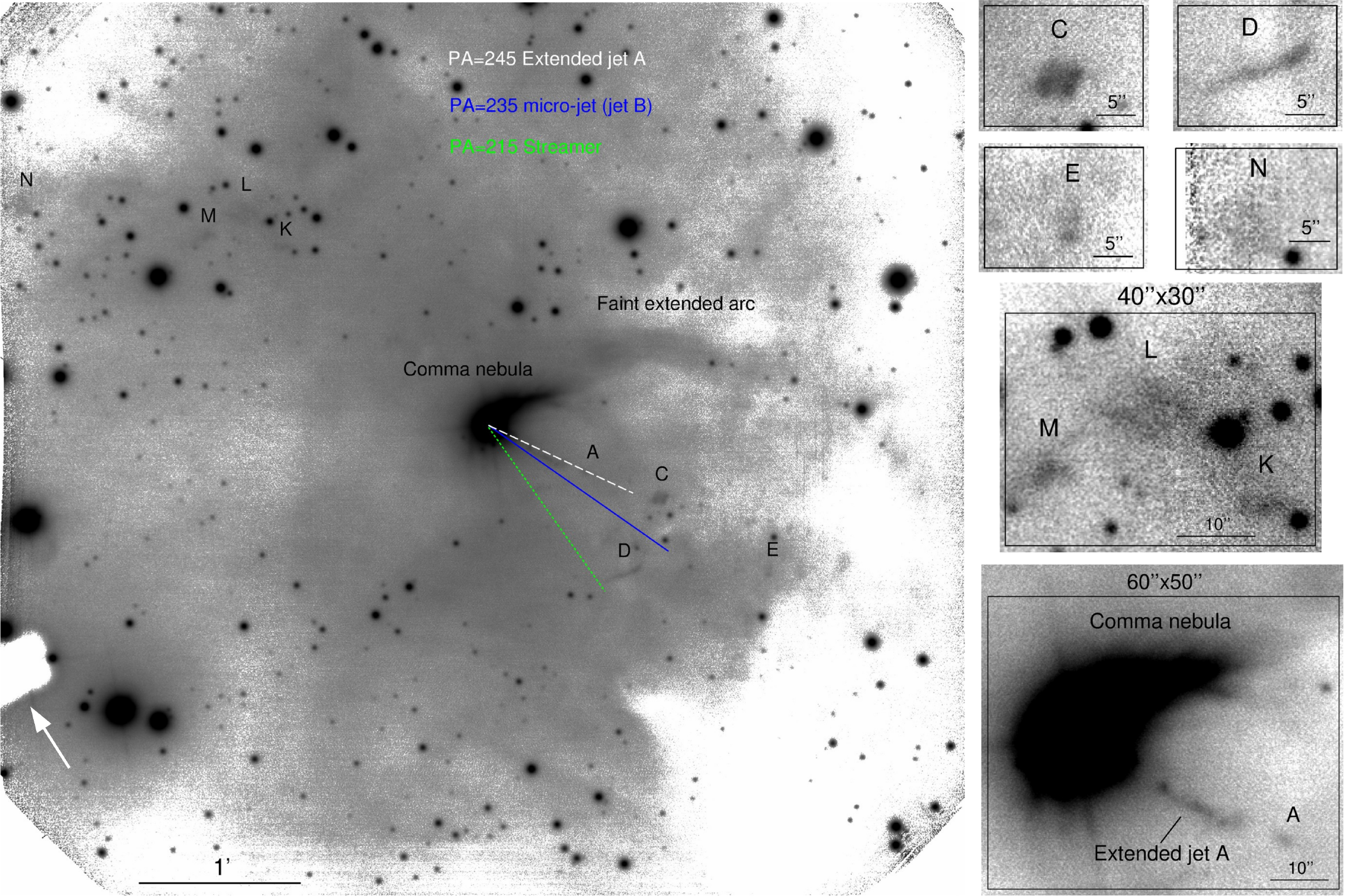}
\caption{{ (\textbf{Left}): }\sii\ image of Z CMa acquired with GMOS attached to Gemini-South. 
The white square shape feature, {indicated with a white arrow,} 
is an artifact from vignetting of the guiding probe. 
{ (\textbf{Right}): } Insets of the resolved features in the \sii\ image. 
The FOVs of the smallest insets are $20''\times15''$ each. 
On all images, north is up, east is to the left{, and the intensity is in log scale to improve the contrast}. See text for more details. }
\label{F-sii_large}
\end{centering}
\end{figure}

The system is surrounded with multiple circumstellar and large-scale outflows. 
The brightest feature around Z CMa is a reflection nebula extending up 
to about $35''$ toward the NW from the central binary. It was discovered 
from photographic plates from 1953 by \citet{1960ApJS....4..337H}. 
On these plates, the reflection nebula has a bar-shaped morphology 
(see Figure~10 in \citet{1960ApJS....4..337H}). 
However, on the newer, higher resolution and more sensitive CCD images, 
the morphology more closely resembles a comma (Figure~{\ref{F-poetzel} left and Figure~}\ref{F-sii_large}). {We refer to this feature as comma nebula.}

In the sub-arcsecond scale, jets emerge from both components---micro-jet A from the primary 
and micro-jet B from the secondary at the position angle (PA, measured from north to east) 245\textdegree\ and 235\textdegree, 
respectively (\citet{2010ApJ...720L.119W}{; Figure~\ref{F-poetzel} right}). Both jets are slightly wiggly 
and show associated knots (\citet{2010ApJ...720L.119W}; \citet{2016A&A...593L..13A}). 
The micro-jet A extends out to about $30''$ and 
is referred to as (extended) jet A {(Figure~\ref{F-poetzel} left)} in the literature. {This }jet was 
discovered by \citet{1989A&A...224L..13P}, and it has also a wiggly nature (\citet{2010ApJ...720L.119W}).

\citet{2002ApJ...580L.167M} discovered another jet-like small-scale feature at PA 215\textdegree. 
This feature is designated in the literature as a streamer, 
and its length is about $2''$ {(Figure~\ref{F-poetzel} right)}. However, it does not emanate from either of the binary 
components. In fact, it appears to start $0''.7$ toward the south (S) from the central binary 
(\citet{2022NatAs...6..331D}). The same authors find a point source a 
further $\sim$$2''$ away from the streamer at the same PA and therefore 
confirm that the streamer is most likely created in a rare flyby event. Furthermore, these authors 
point out that the flyby event explains also the anomalous double-jet activity in 
this system, which considering the masses of the binary components could happen only with 
the probability of less than 1\%. 



Looking at greater distances, further out from the {extended} jet A at the same PA, a 
large-scale outflow was discovered by \citet{1989A&A...224L..13P} 
from the narrow-band \ha\ and \sii\ images acquired in the end of 1990s. 
While the micro and extended jets are primarily detected as one-sided 
objects in the southwest (SW) direction, the large-scale outflow has emission 
features also toward the northeast (NE) ({Figure~\ref{F-poetzel} left}). 
The {large-scale} outflow consists of blobby and elongated features. 
The kinematics of the features refer to a bipolar nature. 

The NE features are all red-shifted, while the SW ones appear blue-shifted. 
Eight features are identified by \citet{1989A&A...224L..13P} in the SW side extending 
up to $4'.7$ and seven features are identified in the NE reaching up to $6'$ from the central object. 
This is the largest known outflow\endnote{Although evolved massive stars can have nebula exceeding this size by far, such as the huge bipolar nebula of the B[e] supergiant MWC 314 extending across 
13 pc (\citet{2008A&A...477..193M}; \citet{2022Galax..10...41L}) as well as 
the 10 pc size elaborate filamentary structures around the Luminous Blue Variable P Cygni 
(see \citet{2006A&A...457L..13B} and references therein).} 
for this type of stars extending across 3.5 pc when considering the distance of 1125 pc \mbox{(\citet{2022NatAs...6..331D}).} 
At the discovery years, the average PA of the {outflow} features was 
60\textdegree\ (equivalent to 240\textdegree). This PA aligns with that of 
the extended \hbox{jet A}, which is associated with the primary, and 
it is therefore widely accepted that the large-scale outflow is a result of the ejections from the 
Herbig Be component. 

The large-scale outflow is what we concentrate on in this paper. 
In particular, we will measure proper motions of the individual features, and in combination 
with their respective radial velocities, we aim to reveal the true 3D nature of this huge nebulosity. 
For that, an accurate distance estimation is essential. 
Several distance estimates for Z CMa exist in the literature. They are all based on the fact 
that Z CMa is a member of the OB association CMa OB1. Published values are 1150 $\pm$ 50 pc 
by \citet{1974A&A....37..229C}, 
990 $\pm$ 50 pc by \citet{2010ApJ...720L.119W}, and more recently 1125 $\pm$ 30 pc 
by \citet{2022NatAs...6..331D}. Throughout this paper, we use the latest 
estimate, 1125 pc, because it was calculated by using the largest number 
of members of the association (50) and is therefore the most accurate one.
We note here that the estimated Gaia Data Release 3 (Gaia DR3) (\citet{2016A&A...595A...1G}; 
\citet{2021A&A...649A...1G}) distance of the \hbox{Z CMa} is not reliable due to the 
very large value of RUWE\endnote{Renormalised Unit Weight Error (RUWE). RUWE is expected to be around 1.0 for sources where the single-star model provides a good fit to the astrometric observations. A value significantly greater than 1.0 (e.g., >1.4) could indicate that the source is non-single or otherwise problematic for the astrometric solution.}, as described in \citet{2022NatAs...6..331D}.


\section{Observations and Data Reduction}\label{sec:obs}

Our first imaging data 
were obtained with the 60-inch telescope at Mt. Palomar on  2002 February 28. 
A single 20-minute exposure in a narrow-band \ha\ filter \hbox{($\lambda=6564.8$ \AA,} $\Delta\lambda=20$ \AA) 
was secured with a seeing of $1''.5$. The field of view (FOV) was $12'.5$,  
and a chosen binning of $2\times2$ provided a pixel scale of $0''.756$ pix$^{-1}$.
This image was reduced using the standard routines in IRAF\endnote{IRAF is distributed by the National 
Optical Astronomy Observatory, which is operated by the Association 
of Universities for Research in Astronomy (AURA) under cooperative 
agreement with the National Science Foundation.} {(\citet{1986SPIE..627..733T,1993ASPC...52..173T})}.

The second set of images of Z CMa was acquired on 2019 September 27 with 
the 8.1 m telescope. {We used the Gemini Multi-Object Spectrographs 
(GMOS, \mbox{\citet{2004PASP..116..425H})} mounted} at Gemini-South as part of the observing proposal AR-2019B-020. 
The images were collected in the narrow band \ha\ G0336 ($\lambda=6567.0$ \AA, $\Delta\lambda=70$ \AA) 
and \sii\ G0335 ($\lambda=6717.2$ \AA, $\Delta\lambda=43$ \AA) filters with the total exposure time of 
145 and 435 seconds, respectively. 
The observations in both filters consisted of several shorter 
exposures that have been dithered to eliminate the 
gaps between the detectors and to minimize contamination (saturation effects) 
due to the bright central star. The bin $2\times2$ was used, yielding a pixel scale of 
$0''.16$ pix$^{-1}$. The FOV of the final reduced images is $6'\times5'.5$. 
The observations were carried out with a seeing between  $1''.3$ and $1''.4$. 
Data reduction was performed using the Gemini software DRAGONS (\citet{2019ASPC..523..321L}). 
{Details of the observations are in Table~\ref{T-obs}.}

\begin{table}[H] 
\caption{{Log of the observations. The first column lists the start 
date of the observing night. Column 2 lists the telescope used. Column 3 contains the 
central wavelength ($\lambda$) and the width ($\Delta\lambda$) of the filter. The last column is the total 
exposure time in seconds.}\label{T-obs}}
\newcolumntype{C}{>{\centering\arraybackslash}X}
\begin{tabularx}{\textwidth}{CCCC}
\toprule
\textbf{Date} & \textbf{Telescope} & \textbf{Filter} & \textbf{Total Exp.}\\
 \textbf{YYYY-MM-DD} &  & \boldmath{$\lambda$/$\Delta\lambda$ (\AA)} & \textbf{Time (s)}\\
\midrule 

2002-02-28 & 60-inch Mt. Palomar & \ha\ 6564.8/20 & 1200\\

2019-09-27 & Gemini-South & \ha\ 6567.0/70 & 145 \\
2019-09-27 & Gemini-South & \sii\ 6717.2/43 & 435 \\
\bottomrule
\end{tabularx}
\end{table}

We also acquired a set of stacked images from the Pan-STARRS images archive (\citet{Waters_2020}) 
that are results of co-adding multiple exposures made between 2010 and 2015 during the
3$\pi$ survey (\citet{Chambers_2016}). 
We downloaded stacked images covering the region 
around Z CMa in $g$, $r$, $i$, and $z$ filters, re-scaled them to the common photometric 
zero point, and created mosaics in each individual filter with the original spatial 
resolution of the Pan-STARRS stacked images of $0.''25$ pix$^{-1}$. We then created 
a composite RGB image from $z$, $i$, and $g$ mosaics with logarithmic intensity scaling applied.
We excluded the $r$ filter from the composite image as it shows the largest number 
of stacking artifacts in the background, and it is mostly unusable for studying the 
morphology of the nebula{r features}. The FOV of the final image was $9'\times9'$.

\subsection{Pre-Analysis Processing of the Narrow-Band Images}\label{S-process}

To accurately analyze the possible morphological and/or kinematical changes 
between our two epochs, our narrow-band \ha\ images first had to be matched pixel by pixel.
For this, we used 32 stars in the FOV whose proper motions were smaller or equal to \hbox{$\pm$5 mas yr$^{-1}$} 
and RUWE < 1.4; all values were taken from the Gaia DR3.  
The 2019 frame was matched against the 2002 frame, because the latter had a larger pixel scale. 
The matching was completed in IRAF using the tasks {\it geomap} and {\it geotran}. 
The errors of the matching were $\sigma_{\mathrm{RA}}=0''.18$ and $\sigma_{\mathrm{DEC}}=0''.23$. 
With this procedure, both frames were given the same pixel scale, $0''.756$ pix$^{-1}$. 
The last step in matching the coordinates is to compensate for the possible 
proper motion of the central star. 
In our case, this effect is insignificant, considering the small proper motion of Z CMa 
(see Section~\ref{sec:f-arc}) and that our two datasets are separated by 17.58 years.
In this stage of the image processing, the frames were ready to be compared 
by blinking to find any obvious movement of the outflow features in the plane 
of the sky or to measure directly the coordinates of individual features to calculate proper motions.

For the features for which the blinking of the frames did not reveal any visual expansion 
and/or which, due to their elongated shape, are not suitable for direct coordinate measuring, 
further processing was needed in order to use the magnification method 
(see Section~\ref{S-D}). These steps included seeing and flux matching. 
The first {was not needed 
because the seeing of the original frames was already similar, and} after pixel by pixel matching, 
{it became equal}.  
Flux matching was completed using the analyzed feature (in our case feature $D$; see Section~\ref{S-D}) 
by summing up all the flux in a rectangle-shaped area equivalent to the size and shape of the feature and 
then arithmetically matching it with the same area flux on the second epoch image. 
Beforehand, the sky was removed. We estimate that the flux matching is accurate 
down to a few percent. 



\section{Results}\label{sec:res}

\begin{figure}[H]
\begin{centering}
\includegraphics[width=13.5cm, trim=0 0 0 0]{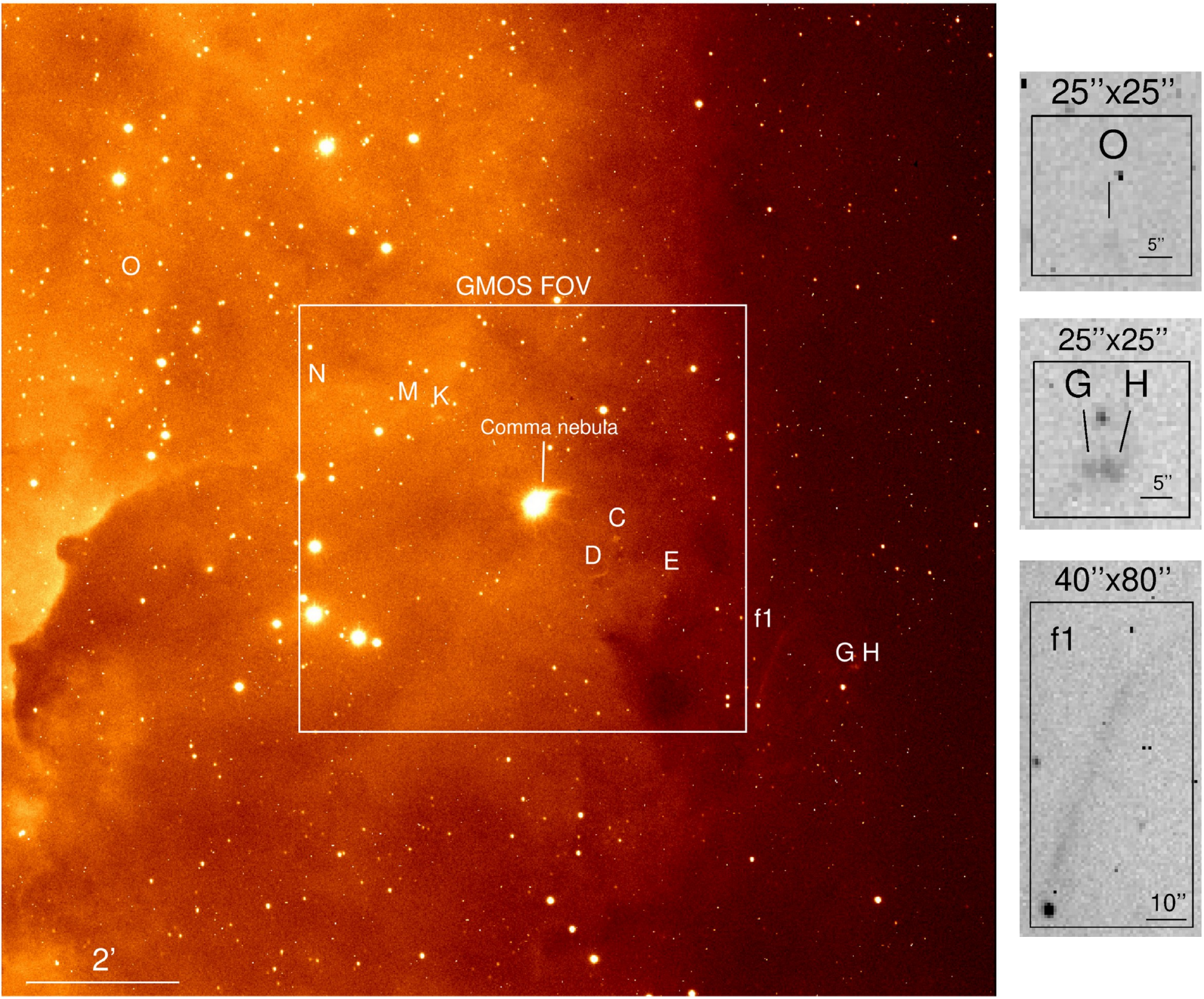}
\caption{{ (\textbf{Left}): } \ha\ image of Z CMa taken in 2002. GMOS FOV is shown for comparison. 
{ (\textbf{Right}): } Insets of the features resolvable in the \ha\ image and which are outside GMOS FOV.
On all images, north is up, east is to the left{, and the intensity is in log scale to improve the contrast. See text for more details.}}
\label{F-Ha2002}
\end{centering}
\end{figure}

In Figure~\ref{F-Ha2002}, we present our 2002 \ha\ image which 
covers the whole large-scale outflow of Z CMa, extending $10'.7$ from NE to SW. 
Our GMOS images from 2019 
have a smaller FOV as demonstrated with the black rectangle in Figure~\ref{F-Ha2002}.  
The GMOS image taken in the lines of \sii\  $\lambda\lambda$6716, 6731 is considerably deeper
and presents the individual 
features with a better S/N (Figure~\ref{F-sii_large}). 
For a meaningful analysis (see Sections~\ref{S-C} and \ref{S-D}), it is important to use 
data in the same filter/spectral lines, especially when the aim is to find
any morphological and/or kinematical changes between two 
epochs. The reason is that the excitation of emission lines from diverse elements can 
occur under different physical conditions so that the lines do not necessarily trace 
the same gaseous regions.
Therefore, we restrict our {proper motion} analysis to the \ha\ frames, because
we do not have a \sii\ frame from 2002. 
In addition, our two \ha\ frames have a similar S/N, hence presenting similar detectability of 
the features, further making them a suitable 
match for the analysis. 
However, we note here that all the features that are resolvable 
in the GMOS \ha\ frame have the same morphology and position in the GMOS \sii\ frame. 

We refer to the individual features as they have been named by previous authors. 
The features of the {large-scale} outflow were named by \citet{1989A&A...224L..13P} 
using capital letters from $A$ to $O$ ({Figure~\ref{F-poetzel}}).
In addition, the designation 
of $f1$ and $f2$ was given to refer to the filaments in the SW side nearby the 
blobby features $F$, $G$, and $H$. On our figures, the labels of the features are always directly 
above the feature itself, apart from the label $f1$ which is to the left from 
the feature (Figure~\ref{F-Ha2002}). 
We note here that not all the features presented by \citet{1989A&A...224L..13P} are 
detectable and/or resolvable in our 2002 \ha\ frame due to the slightly lower S/N. 
From our GMOS \sii\ frame, we could identify the features $A$, $B$, $C$, $D$, $E$, $M$, $N$, $K$, and $L$. 
Features $O$, $G$, $H$, and $f1$ are outside the GMOS FOV.  
We could not identify features $I$ and $J$, situated between the central star and the feature $K$, 
on any of our images, which is probably due to a slightly lower S/N of our images 
compared to the images from \citet{1989A&A...224L..13P}.

The morphology of the large-scale {outflow} has not changed during the past 30 years when comparing 
our 2019 image with the 2002 one and the one from 1988--1989 from \citet{1989A&A...224L..13P} 
(compare their Figure 1 with our Figures~\ref{F-sii_large} and ~\ref{F-Ha2002}). 
The large-scale outflow has a bipolar nature and it consists of individual features 
(features $A$ to $O$) with varying shapes---blobby, elongated, filamentary, arced.  
The PA of the outflow is 
$\sim$60\textdegree\ (or $\sim$240\textdegree), as measured from our images. 
The approximate value is due to the slightly different PAs of individual features. 
Nevertheless, this shows that the PA of the outflow has not changed during the past 
30 years either ($\sim$60\textdegree\ is measured also by \citet{1989A&A...224L..13P}).

In our \sii\ frame (Figure~\ref{F-sii_large}), we refer to a few other features 
related to the ejections from Z CMa: in particular, 
the extended jet A (see also Figure 3 in \citet{2010ApJ...720L.119W}), 
the PA of the micro-jet B (see also Figure 1 in \citet{2016A&A...593L..13A}), 
and the PA of the streamer (see also Figure 2 in \citet{2015A&A...578L...1C}). 
Figure~\ref{F-sii_large} shows also the previously known bright 
comma nebula which is almost perpendicular to the large-scale outflow.
Furthermore, our image reveals 
another, fainter and previously not detected extended arc-shaped
feature in the NW direction, which will be discussed further in Section~\ref{sec:f-arc} and 
\ref{sec:dis}.

As a first step in finding any expansion in the plane of the sky, we have 
used the simple blinking of the two \ha\ matched frames. 
It reveals that feature $C$ has a visible expansion, 
while the rest of the features appear to stand still. 
Overall, a reliable analysis is only possible for the brightest 
features, which are those labeled with $C$ and $D$. Therefore, we focus in the following on these 
two and determine their proper motions before we take a closer look at the faint arc structures.

\subsection{Proper Motion Calculations of Feature $C$}\label{S-C}

Feature $C$ is one of the brightest among the large-scale outflow. It has a roundish shape 
and is clearly detectable in both of our \ha\ 
frames taken in 2002 and 2019. The exact temporal separation of the two images is 17.58 years. 
The shape of feature $C$ does not change during that period. 
As mentioned above, feature $C$ presents the fastest motion from the central 
star compared to other features. In addition, feature $C$ also has the   
largest radial velocity compared to other features as measured by \citet{1989A&A...224L..13P}. Therefore, 
considering the inclination angle out of the plane of the sky (see below), 
it is not surprising that this feature 
would show a clear expansion in the plane of the sky while others do not.
The movement in the plane of the sky of feature $C$ is in accordance with the general 
direction of the features in SW direction, confirming that it must have been ejected from Z CMa.

Due to the roundish shape of feature $C$, it was possible to measure directly its 
central coordinates on both of our images. 
The total movement in the plane of the sky during the 17.58 years considered is
 $1''.4$, yielding 
 a proper motion of $0''.08$ yr$^{-1}$ and a tangential velocity of 
$\sim$$420$ km s$^{-1}$. 
Considering the radial velocity of $-$390 km s$^{-1}$ of feature $C$ (\citet{1989A&A...224L..13P}), its expansion velocity 
is about $580$ km s$^{-1}$ and the inclination out of the plane of the sky is 
43\textdegree\ using the ordinary cosine relation between the velocity vectors ($ i = arccos ( v_{sky}/v_{exp})$).
The found inclination angle agrees with the estimates made for the micro-jet B, 
which was proposed to have an inclination angle between 28\textdegree\ and 64\textdegree\ 
(\citet{2016A&A...593L..13A}) according to the 
tangential and radial velocity estimates by \citet{2010ApJ...720L.119W}. 

The distance of feature $C$ from the 
central star is $68''$ and $69''$ during the observations taken in 2002 and 2019, respectively. 
The position angle of feature $C$ has not changed during the time duration 
between our 2002 and 2019 images, and it is 246\textdegree. 
Precise measurements with errors for all the 
calculated values are in Table~\ref{T-propm}.



\begin{table}[H] 
\caption{Results of the calculations from direct measuring of feature $C$ and from using the magnification 
method for feature $D$.  A distance of 1125 pc  toward Z CMa was adopted. See text for more details. \label{T-propm}}
\newcolumntype{C}{>{\centering\arraybackslash}X}
\begin{tabularx}{\textwidth}{Ccc}
\toprule
\textbf{Description} &	\textbf{Feature} \boldmath{$C$} & \textbf{Feature} \boldmath{$D$}\\
\midrule
Total movement in the plane of the sky ($''$)	 & $1.4\pm0.3$ & ---\\
Proper motion $\mu$  ($''$ yr$^{-1}$) & $0.08\pm0.02$& 0.013 $\pm$ 0.004\\
Radial velocity $^{a}$ $v_{rad}$ (km s$^{-1}$) & $-$390 $\pm$ 24 &$-$110 $\pm$ 24\\
Tangential velocity $v_{sky}$ (km s$^{-1}$) & 423 $\pm$ 88 & 69 $\pm$ 23\\
Expansion velocity $v_{exp}$ (km s$^{-1}$) & 576 $\pm$ 67& 130 $\pm$ 24\\
Inclination out of the plane of the sky $i$ (\textdegree) & 43 $\pm$ 15& 58$\pm$14\\
Distance from the central star in 2002 $d_{2002}$ ($''$) & 67.8 $\pm$ 0.3& 75.3 $\pm$ 0.8\\
Distance from the central star in 2019 $d_{2019}$ ($''$) & 69.2 $\pm$ 0.3& 75.5 $^{b}$ $\pm$ 0.8\\
PA at 2002 (\textdegree) & 246.5 $\pm$ 0.2& 222.9 $\pm$ 0.6\\
PA at 2019 (\textdegree) & 246.6 $\pm$ 0.2& 222.9 $\pm$ 0.6\\
Age at 2002 (years) & 854 $\pm$ 177& 5859 $\pm$ 1953\\
Magnification factor  $M$ & --- & 1.003 $\pm$ 0.001\\
\bottomrule
\end{tabularx}
\noindent{\footnotesize{
$^{a}$ From \citet{1989A&A...224L..13P}. 
$^{b}$ Calculated from $d_{2002}$ and our derived proper motion.}}
\end{table}

Using the above calculated proper motion, the distance from the central star, and 
assuming constant expansion velocity since the ejection, it is possible to 
calculate the age of feature $C$ at our first epoch, 2002. It is 
on the order of 850 years, which is in accordance with the estimates 
in \citet{1989A&A...224L..13P}.


\subsection{Proper Motion Calculations of Feature $D$}\label{S-D}


The second feature for which we were able to calculate the expansion in the plane of the sky is 
the arc-shaped feature $D$. 
Due to its elongated shape, a direct measuring of the coordinates was not an appropriate 
method. For that reason the magnification method (see, e.g., 
\citet{2007A&A...465..481S}; \citet{2018A&A...612A.118L}) was used, which is suitable to find the proper motion 
of extended structures without a clear central point and/or when the total movement in the 
plane of the sky is as small as a tenth of a pixel. Both criteria are valid 
for feature $D$.
The magnification method is based on finding a magnification factor $M$, which represents an 
image with minimum residuals of the magnified 
first epoch image, which is subtracted from the second epoch image.  
The method provides the proper motion, tangential velocity, and age. 
In order to use the magnification method, the frames being analyzed have to have  
coordinates, seeing, and flux matched. This was completed using the procedures described in 
Section~\ref{S-process}. Further details about the 
magnification method and the derivation of the formulas used in the following can 
be found in Section 3.3 of the PhD thesis by \citet{2019arXiv191004157L}.
The best magnification factor for feature $D$ was determined to be $M = 1.003 \pm 0.001$. 
However, we note here that this result should be used with caution. We can say with confidence that 
$M$ is not larger than $1.003$. Consequently, all the following numerical values should 
be taken as upper limits. The proper motion can be calculated, in convenient units, in the following way 

\begin{equation}\label{eq-pm}
\mu ['' yr^{-1}]= \frac{ (M-1) \cdot d [''] } {\Delta t [yr] },
\end{equation}
where $d$ is the distance of feature $D$ from the central star on the first epoch, in our case the year 2002, 
and $\Delta$t is the time interval 
between the two epochs. In the case of the elongated feature $D$, the distance from the central star 
is somewhat challenging to estimate. However, we are confident that when considering 
a somewhat larger error of 1 pixel, 
it is accurate enough to serve the purpose of the simple calculations presented in this paper.  
Hence, the distance of feature $D$ from the central star at our 2002 epoch is $75''\pm1''$, 
and considering Equation~(\ref{eq-pm}), the proper motion becomes $\mu = 0''.013$ yr$^{-1}$. 
As for feature $C$, 
the precise values with their errors for all the calculations 
for feature $D$ are in Table~\ref{T-propm}.

Using the proper motion 
and the distance to feature $D$, it is possible to calculate the tangential velocity in the following 
convenient units

\begin{equation}\label{e-vsky}
v_{sky}\ [km\ s^{-1}] = 4.74 \cdot \mu\ [''\ yr^{-1}] \cdot D \ [pc].
\end{equation}

We consider the distance to Z CMa to be 1125 pc and therefore $v_{sky} = 69$ km s$^{-1}$. 
The radial velocity of feature $D$ has been measured to be $-$110 km s$^{-1}$ 
(\citet{1989A&A...224L..13P}), 
which results in an expansion velocity of \hbox{130 km s$^{-1}$}. The inclination angle would therefore 
be slightly larger than for feature $C$ but with a value of 58\textdegree\, again matching with the estimates 
in \citet{2016A&A...593L..13A}.

The magnification factor can also be used to calculate the age of the feature at the first epoch, 

\begin{equation}\label{e-age2}
T\ [yr]= \frac{ \Delta t \ [yr] } {(M-1)}, 
\end{equation}
which for feature $D$ gives a value of about 6000 yrs.

The PA of feature $D$ is constant between our two observing epochs 
and has a value \mbox{of 223\textdegree.} 

\subsection{Proper Motion of Other Features}\label{sec:others}

We tried to measure the expansion in the plane of the sky for the two other 
features, $E$ and $K$, which were considerably fainter than the features $C$ and $D$ 
but still resolvable compared to the rather marginal detections of the features $L$, $M$, and $N$. 
Features $E$ and $K$ have an irregular shape, and therefore,   
the magnification method was used. We find no measurable 
movement in the plane of the sky for both features. 
For feature $E$, it is somewhat expected due to its RV 
being 0 km s$^{-1}$ (\citet{1989A&A...224L..13P}), while the RV of feature $K$ 
is quoted as +55 km s$^{-1}$ by the same authors. 
Considering the pixel scale of $0''.756$ of our matched 
images and the fact that the magnification method is able to measure an expansion 
of about one tenth of a pixel, the smallest tangential velocity that we should be able to detect 
is $\sim$20 \kms. This, in return, would 
mean an inclination out of the plane of the sky 70\textdegree, 
which, within our error estimate of $\pm$10\textdegree, 
agrees with the estimates by \mbox{\citet{2016A&A...593L..13A}.}

\subsection{Faint Extended Arc}\label{sec:f-arc}



\begin{figure}[H]
\begin{centering}
\includegraphics[width=13.5cm, trim=0 0 0 0]{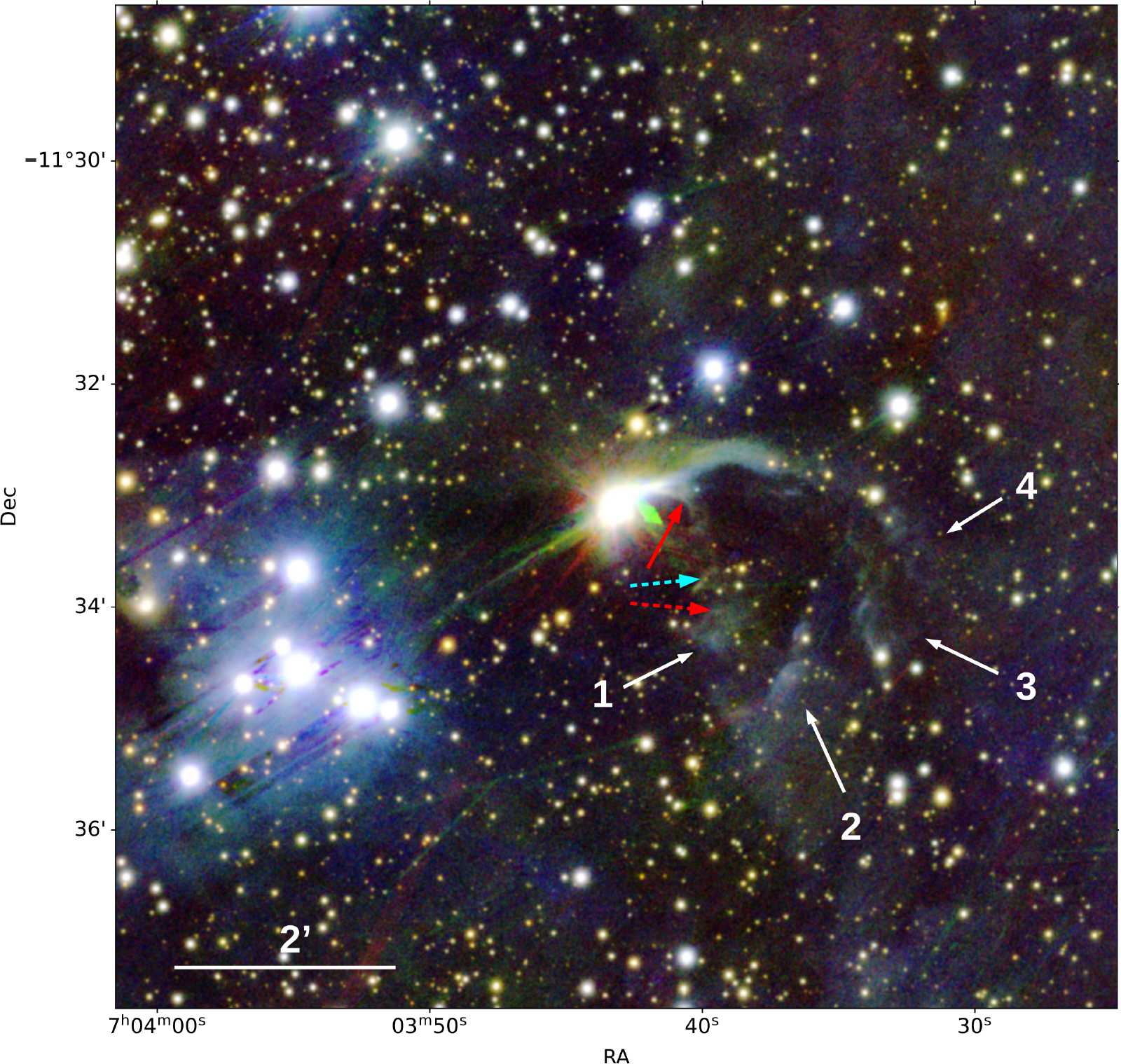}
\caption{Pan-STARRS RGB image presenting the faint extended features around Z CMa. 
Red channel corresponds to $z$ filter, green channel corresponds to $i$, and blue channel corresponds to $g$ filter. 
{The intensity is in log scale to improve the contrast.}
North is up and east is to the left. FOV $9'\times9'$.  See text for more details.}
\label{F-pan}
\end{centering}
\end{figure}

From our deep GMOS \sii\ image, we have discovered that the bright comma nebula has a fainter continuation. We designate this 
feature a faint extended arc (see Figure~\ref{F-sii_large}). 
Despite the lower S/N of our GMOS \ha\ image, the faint arc is also detectable on that frame,  
but we refrain from showing it because it does not provide new information. 
With an orientation of the faint arc  toward the NW, it is perpendicular to 
the main large-scale outflow. 
The arc is more pronounced on the Pan-STARRS RGB image (Figure~\ref{F-pan}) 
on which additional related features become visible. 
We detect a repeating pattern of filaments, inside the main arc, which seem to 
mimic “feathers” (see white arrows in Figure~\ref{F-pan}). 
Interestingly, while most of the feathers do not have a direct 
connection to the star, the closest feather to the central binary, designated with number 1, 
seems to have connecting filaments. These filaments start from the 
bright comma nebula as a small arc resembling a “fishtail” (marked 
with solid red arrow) and then continue  toward feather 1 with 
less homogeneous emission. From the image, it is visible that 
the diffuse emission in the direction of feather 2 is continuous  
further than that of our Pan-STARRS FOV. However, the faint extended arc which ends with feather 4 
extends up to $3'$ toward the west.

\section{Discussion}\label{sec:dis}


Looking at the Figure~\ref{F-pan}, it is 
tempting to assume that the newly discovered faint extended features, 
as well as the bright comma-shaped feature, are created by matter 
expelled from the central star during a single or multiple independent mass ejection 
event(s) and now appear misplaced from the central position of the object 
due to the 
movement of Z CMa through the interstellar medium. However, 
the proper motion of the star, $\mu_{RA}=-4.4$ mas yr$^{-1}$ and $\mu_{DEC}=2.0$ mas yr$^{-1}$ (Hipparcos-2 catalog, 
\mbox{\citet{2007A&A...474..653V}),} does not support that. Z CMa is 
moving  toward the faint extended arc and not away from it. 
Our choice of using the proper motion values from the Hipparcos-2 catalog is 
due to the mentioned large RUWE value of Z CMa in Gaia DR3 (see Section~\ref{sec:intro}),  
referring to a problematic astrometric solution. The latter is most probably a result 
of the binary nature of the object and its generally rather bright photometric values. 
However, we note that Gaia DR3 proper motion values agree relatively well with the Hipparcos-2 
catalog.  

Another possibility, due to the roundish shape, is that the faint extended arc with the feathers 
is related to the orbital motion of the binary. Considering 
the recent high-contrast imaging polarimetry observations (\citet{2015A&A...578L...1C}),  
the orbital movement of the FU Ori companion around the primary is 0\textdegree.7 yr$^{-1}$ 
when considering a circular orbit. This implies an orbital period of about 500 years. 
Currently, the FU Ori companion is located in the southeast (SE) direction
from the primary (e.g., Figure 2 in \citet{2017A&A...597A..91B}). Therefore, considering 
the hypothesis of a single ejection and 
the position of the faint extended arc and its feathers, they could have been 
ejected half a period ago. However, taking into account the distance  
of these features from the central binary, it would imply a peculiarly large 
tangential velocity of $\sim$3700 \kms, which has not been measured in any other features related to 
the small- or large-scale outflows of Z CMa. The obvious repeating pattern of the feathers 
could also suggest consecutive mass ejections occurring on every orbit at a 
specific location. We can then assume that the closest feather could have been ejected 
half a period ago and the next ones could have been ejected 1.5, 2.5, and 3.5 periods ago, respectively. 
The equivalent tangential velocities would then be about 1800, 900, 600, and 
530 \kms, considering distance estimates for each feather from the central binary 
of 85, 130, 148, and 180$''$, respectively. While the radial velocities of $-$600 \kms\ have 
been measured to be associated with the micro-jets 
(\citet{1989A&A...224L..13P}; \citet{2010ApJ...720L.119W}), all these estimated 
values exceed our measured tangential velocities of the large-scale 
outflow features (see Table~\ref{T-propm}). If we consider our largest tangential 
velocity of 420 \kms, and the largest extent of the faint arc structures (3$'$), 
it would have had to be ejected $\sim$2300 years ago. This timescale is comparable 
with the ages we have found for the large-scale outflow features $C$ and $D$, 850 and 5900 
years, respectively. 
However, if these newly discovered features are related to the mass ejections occurring 
at the particular location during the orbit of the FU Ori-type companion around the primary, 
it is more likely that the 
orbit is not circular but elliptical. The latter would include periastron passage, which 
can enhance the mass-loss and possibly initiate outflows, as has been seen  
in other binary systems (e.g., in a case of symbiotic binary R Aquarii  (\citet{2021gacv.workE..41L}) and proposed for the formation of the circumbinary molecular ring in the B[e] supergiant system GG Car 
(\citet{2013A&A...549A..28K})). 
At the same time, while feathers 2 and 3 do not seem to be physically connected 
with the central binary, the faint extended arc and its extension feather 4 are clearly a continuation 
from the star and its bright comma nebula,  
implying a constant flow of matter. At this point, we 
also mention that when inspecting the Figure~\ref{F-pan} more thoroughly, 
it is possible that there is a dark cloud 
blocking the connections of the feathers with the central binary or with the faint 
arc, as there 
are no stars detected in the west from the binary below the faint 
arc. However, according to the study based on 2 Micron All Sky Survey by 
\citet{2011PASJ...63S...1D}, there are no dark clouds in the FOV of our Figure~\ref{F-pan}. 

We can further calculate the possible tangential velocities related 
to the new discovered features 
when considering their maximum extent of $3'$ and 
the ages found for the features $C$ and $D$. 
The age of 850 years would result in a velocity of about 1100 \kms, 
while the 5900 years (feature $D$) would result in a velocity of about 160 \kms. The latter is more in line with the 
tangential velocities measured in this work and with the radial velocities 
measured by \citet{1989A&A...224L..13P}. It is possible that the 
elongated feature $D$, which indeed has a larger deviation from the average PA of the large-scale 
outflow compared to other features, potentially referring to a different 
origin, is related to the feather 1. However, the seemingly similarly shaped 
feature marked with dashed cyan arrow in Figure~\ref{F-pan} 
is not feature $D$. Feature $D$ is 15$''$ further away from the central star 
and has a slightly smaller PA. A red dashed arrow is indicating its position in 
Figure~\ref{F-pan}, and it shows that this feature is situated on 
the edge of the feather 1, but there is no brightness 
enhancement in that position other than the boarder line of the feathery feature. 

On the other hand, it cannot be ruled out that the faint features are accidentally 
aligned with Z CMa and that they are actually part of the huge nebula Sh 2-296
on which edge Z CMa is located (see Figure 6 in \citet{2019A&A...628A..44F}).
However, due to the obvious positional proximity to Z CMa, we are inclined 
to favor the idea that the discovered features are connected to Z CMa rather than 
being aligned by chance.


The new detected faint features as well as our measured 
different ages of different features (850 versus 5900 years for feature $C$ and $D$, respectively) 
are in accordance with the nature of the Z CMa binary, which 
has had several eruptions in the past. The knotty 
nature of the {large-scale} outflow as well as the (micro) jets 
(e.g., \mbox{\citet{2010ApJ...720L.119W},} \mbox{\citet{2016A&A...593L..13A})} additionally refer 
to several discrete mass ejections. In addition, the RVs of the individual features vary a  
lot with values reaching up to $\sim$$\pm400$ $\kms$ \mbox{(\citet{1989A&A...224L..13P}), }
further supporting the scenario 
that the central object has experienced in the past several mass ejections with  
different initial velocities.

{Another possible explanation could be that the measured tangential velocities 
of feature $C$ and $D$ have not been constant since their ejection from the central binary, 
which, in return, would affect the calculated ages. However, observationally, we cannot assess it 
at this point. We have not found suitable observational data prior to 2002 to check the potential
velocity canges before our first dataset. In addition, considering the measured velocities, we will have 
to wait for at least another $\sim$20 years to obtain a new set of observations, which could 
potentially show a change in tangential velocities. 
However, independently of the two scenarios, we wish to emphasize that our measured ages 
are more precise than previous estimates 
(\citet{1989A&A...224L..13P}) because we have accurately measured the tangential velocities, 
while formerly, those were approximated according to the radial velocities and 
the possible inclination angle of the large-scale outflow.}

As measured from our 2002 and 2019 images, the PA of feature $D$ is 223\textdegree\ 
and does not change during our observing period, 17.58 years. To expand the 
epoch of analsyis, 
we estimate from the schematic Figure 1 in \citet{1989A&A...224L..13P} (they do not 
publish any numerical values of the PAs nor the distances from the central star) 
that the PA of feature $D$ at 
their observing time between 1989 and 1990 was $\sim$225\textdegree.  
Therefore, we conclude 
that the PA of feature $D$ has not changed during the past 30 years. Following this 
result, the claim 
made by \citet{2010ApJ...720L.119W} that feature $D$ is related with the 
micro-jet B (emanating from the FU Ori-type companion with a PA 
$\sim$235\textdegree, which is indicated with the blue solid line in Figure~\ref{F-sii_large}), 
according to their observations, is not supported by 
our precise PA measurements.

We investigate further whether feature $D$ could be related to the 
additional sub-arcsecond component emerging from Z CMa, the jet-like structure 
identified as a streamer 
(see Figure 3 in \citet{2002ApJ...580L.167M}, Figure 2 in \citet{2015A&A...578L...1C}, 
Figure 1 in \citet{2016SciA....2E0875L}, and Figure 1 in \citet{2022NatAs...6..331D}). 
Unfortunately, these authors do not provide any PA measurements for the straight part of the 
streamer (the outer edge of this feature is slightly curved  toward west),  
but from their figures, we can estimate it to be approximately 215\textdegree. 
We also note that the streamer does not start from the central binary 
but about $0''.7$ straight  toward the south from the binary (see Figure 2 in \citet{2015A&A...578L...1C}). 
We indicate the PA of this feature with the green dotted line in our Figure~\ref{F-sii_large}.
Even though the angle of the streamer is more similar to the PA of feature $D$ 
than the angle of the micro-jet B is, it is still clearly evident that their PAs do not align. 
\citet{2022NatAs...6..331D} explains the streamer as a result of the 
flyby event because they discovered a faint component whose PA matches with the one 
of the streamer. The fact that feature $D$ is not aligned with the streamer 
provides an additional support for the flyby event because it shows that 
the streamer is not related to any small or large-scale ejecta. 


\section{Conclusions}

We have presented the first proper motion study of features $C$ and $D$ within 
the large-scale outflow of Z CMa.
The two very different proper motion values obtained for these two 
features confirm the previous 
suggestion that the large-scale outflow is a result of several active ejection phases 
with varying initial velocities in the life of Z CMa. Our precise position angle 
measurements of the same features reveal that they are not aligned with the 
feature streamer, providing further support for the occurrence of the flyby event in this 
complex system.

We have discovered new features most probably related to Z CMa---a faint extended arc 
with several features mimicking feathers. 
It is very likely that these features are connected to the central binary and 
are the result of previous mass ejection(s) possibly related to 
the orbital motion of the binary system.




\newpage
\authorcontributions{
Conceptualization and Project administration, T.L., M.K., L.C.; 
Resources and Data curation: T.L., L.C., S.K. and A.M.;
Formal analysis, Investigation, Software, Methodology and Validation, T.L.; 
Supervision, M.K.;
Visualization, T.L., S.K.; 
Writing---original draft preparation, T.L., S.K.; 
Writing---review and editing, T.L., M.K., S.K., L.C., A.M.; 
Funding acquisition, M.K., L.C., S.K.
All authors have read and agreed to the published version of the manuscript.
}

\funding{This research was funded by the Czech Science Foundation (GA \v{C}R, grant number \hbox{20-00150S}), by CONICET 
(PIP 1337) and the Universidad Nacional de La Plata (Programa de Incentivos 11/G160).
The Astronomical Institute of the Czech Academy of Sciences is supported by the project RVO:67985815. This project has also received funding from the European Union's Framework Programme for Research and Innovation Horizon 2020 (2014--2020) under the Marie Sk\l{}odowska-Curie Grant Agreement No. 823734.
S.K. acknowledges support from the European Structural and Investment Fund
and the Czech Ministry of Education, Youth and Sports (Project CoGraDS-CZ.02.1.01/0.0/0.0/15\_003/0000437).}

\dataavailability{Publicly available GMOS-S and Pan-STARRS datasets analyzed in this study 
can be retrieved from dedicated archives \url{https://archive.gemini.edu/searchform} (accessed on 27 February 2023), and 
\url{https://ps1images.stsci.edu/cgi-bin/ps1cutouts} (accessed on 27 February 2023), 
 respectively. The Mt. Palomar image is available 
from the corresponding author on reasonable request.}

\acknowledgments{
This research made use of the NASA Astrophysics Data System (ADS) and of
the SIMBAD database, which is operated at CDS, Strasbourg, France. 
This publication is based on observations obtained at the international Gemini Observatory, which is a program of NSF’s NOIRLab 
(processed using DRAGONS (Data Reduction for Astronomy from Gemini Observatory North and South)), which is managed by the Association of Universities for Research in Astronomy (AURA) under a cooperative agreement with the National Science Foundation on behalf of the Gemini Observatory partnership: the National Science Foundation (United States), National Research Council (Canada), Agencia Nacional de Investigaci\'{o}n y Desarrollo (Chile), Ministerio de Ciencia, Tecnolog\'{i}a e Innovaci\'{o}n (Argentina), Minist\'{e}rio da Ci\^{e}ncia, Tecnologia, Inova\c{c}\~{o}es e Comunica\c{c}\~{o}es (Brazil), and Korea Astronomy and Space Science Institute (Republic of Korea) under program ID GS-2019B-Q-210. 
The Pan-STARRS1 Surveys (PS1) and the PS1 public science archive have been made possible through contributions by the Institute for Astronomy, the University of Hawaii, the Pan-STARRS Project Office, the Max--Planck Society and its participating institutes, the Max Planck Institute for Astronomy, Heidelberg and the Max Planck Institute for Extraterrestrial Physics, Garching, The Johns Hopkins University, Durham University, the University of Edinburgh, the Queen's University Belfast, the Harvard--Smithsonian Center for Astrophysics, the Las Cumbres Observatory Global Telescope Network Incorporated, the National Central University of Taiwan, the Space Telescope Science Institute, the National Aeronautics and Space Administration under Grant No. NNX08AR22G issued through the Planetary Science Division of the NASA Science Mission Directorate, the National Science Foundation Grant No. AST-1238877, the University of Maryland, Eotvos Lorand University (ELTE), the Los Alamos National Laboratory, and the Gordon and Betty Moore Foundation.
This work has made use of data from the European Space Agency (ESA) mission
{\it Gaia} (\url{https://www.cosmos.esa.int/gaia}), processed by the {\it Gaia}
Data Processing and Analysis Consortium (DPAC,
\url{https://www.cosmos.esa.int/web/gaia/dpac/consortium}) (accessed on 27 February 2023). Funding for the DPAC
has been provided by national institutions, in particular the institutions
participating in the {\it Gaia} Multilateral Agreement.}

\conflictsofinterest{The authors declare no conflict of interest.} 



\abbreviations{Abbreviations}{
The following abbreviations are used in this manuscript:\\

\noindent 
\begin{tabular}{@{}ll}
Z CMa & Z Canis Majors \\
FU Ori & FU Orionis \\
PA & position angle \\
FOV & field of view \\\end{tabular}

\noindent 
\begin{tabular}{@{}ll}
RUWE & renormalized unit weight error \\
S & south \\
NW & northwest \\
NE & northeast \\
SW & southwest \\
SE & southeast \\ 

\end{tabular}
}

\appendixtitles{no} 
\appendixstart
\appendix

\begin{adjustwidth}{-\extralength}{0cm}
\printendnotes[custom] 

\reftitle{References}

\PublishersNote{}
\end{adjustwidth}
\end{document}